\documentclass[12pt, letterpaper]{article} 
\usepackage{natbib, hyperref, graphicx}
\usepackage{geometry}
\usepackage{amssymb}
\usepackage{verbatim}
\usepackage{subfigure,wrapfig}
\usepackage{enumitem}
\setlist[enumerate]{itemsep=0mm}
\usepackage{float}
\usepackage{lscape}
\usepackage{lineno}

\usepackage[font=small]{caption}

\newcommand{\eyja}{Eyjafjallaj\"okull}

\title{The development of near-vent volcanic ash cloud layers due to
  inhomogeneous atmospheric turbulence and relationship to wind shear}

\author{Marcus Bursik $^{1,}$*{ORCID: 0000-0002-9312-5202}, Qingyuan Yang
  $^{2}$ \\ Adele Bear-Crozier $^{3}$, Michael Pavolonis
  $^{4}$ and Andrew Tupper $^{3}$ \\
$^{1}$ \quad Center for Geohazards Studies, \\ University at Buffalo,
Buffalo NY USA\\
$^{2}$ \quad Earth Observatory of Singapore, \\ Nanyang Technological
University, Singapore\\
$^3$ \quad Bureau of Meteorology, Melbourne, Australia\\
$^4$ \quad NOAA Cooperative Institute for \\ Meteorological Satellite Studies
University of Wisconsin, Madison WI USA \\
Correspondence: mib@buffalo.edu; Tel.: +1-716-645-4265}

\begin{document}

\maketitle

\abstract{Volcanic ash clouds often become multilayered and thin
  with distance from the vent.  We explore one mechanism for
  development of this layered structure.  We review data on the
  characteristics of turbulence layering in the free atmosphere, as
  well as examples of observations of layered clouds both near-vent and
  distally.  We then explore and contrast the output of volcanic ash
  transport and dispersal models with models that explicitly use the
  observed layered structure of atmospheric turbulence.  The results
  suggest that the alternation of turbulent and quiescent atmospheric
  layers provides one mechanism for development of multilayered ash
  clouds by modulating the manner in which settling occurs.}


\section{Introduction}

Volcanic ash is a multi-billion dollar economic hazard to aviation, as
shown during the 2010 eruptions of Eyjafjallaj{\"o}kull, Iceland
\citep{c94, mazzocchi20102010}. It is also a risk to flight safety,
with hundreds of encounters of varying severity recorded, and several
instances of multiple engine flame-out in flight.  The International
Airways Volcano Watch (IAVW), which seeks to safely separate aircraft
from volcanic ash in flight, relies on detecting areas of ash and
forecasting its future movement \citep{tupper2007facing}.  However,
the forecasting of ash presence and concentration is generally poorly
resolved vertically, although there is some progress in this
direction, e.g., \citep{HEINOLD2012195,
  kristiansen2015stratospheric}. Aircraft flying in a supposedly
ash-contaminated region at a particular altitude may encounter no ash
or significant and potentially damaging amounts, due to the high
degree of ash stratification with altitude. The improved understanding
and forecasting of stratification would assist enormously in managing
the hazard and support the continuing development of the IAVW.

\begin{figure}[!htb]
  \centering
  \includegraphics[width=0.7\textwidth]{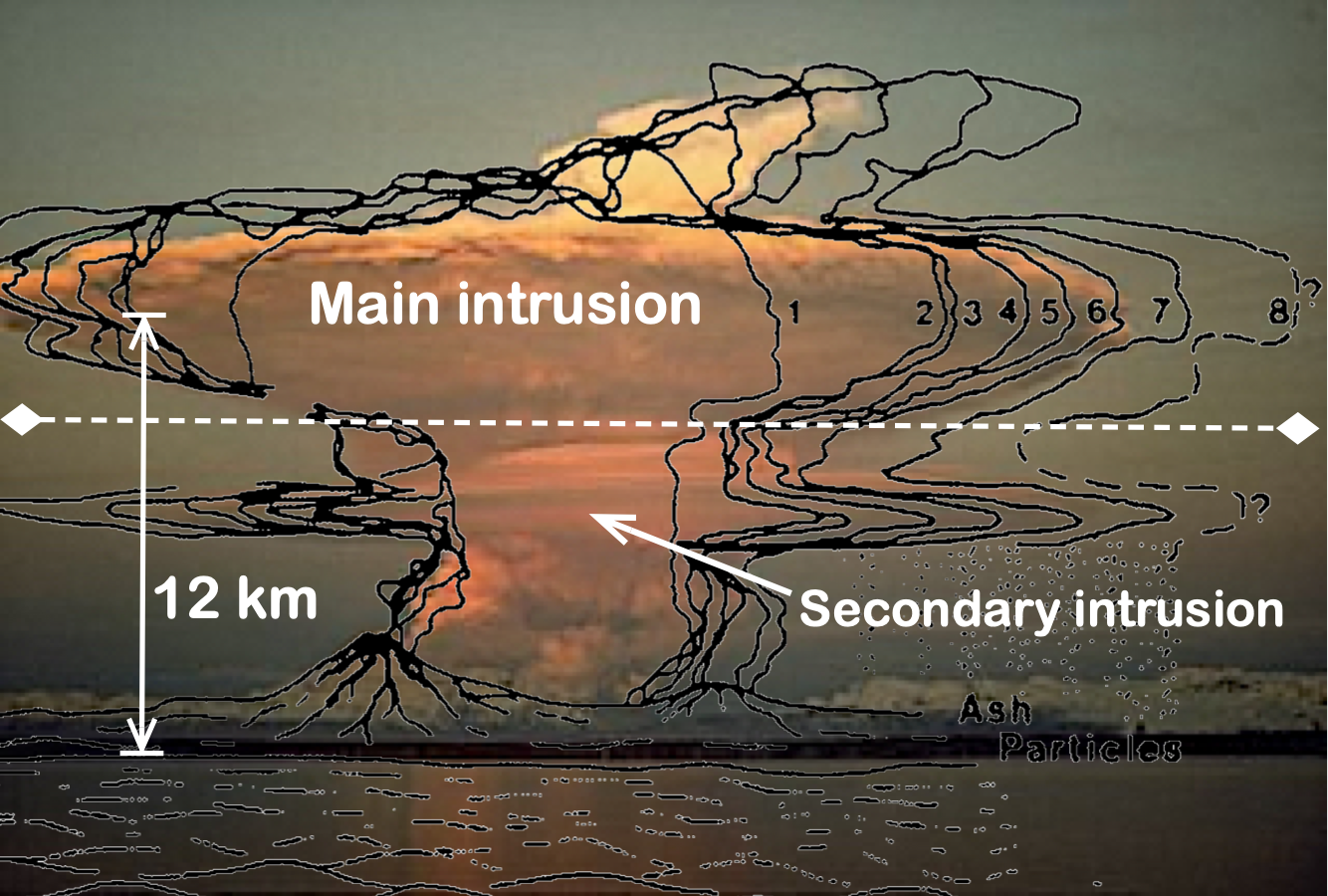}
  \caption{Photograph and tracing of eruption of Redoubt volcano, AK,
    21 Apr 1990, showing plume overshoot, main umbrella intrusion and
    particle rich secondary intrusion.  Tracings show growth of cloud
    layers through different time steps (numbered solid lines), long
    dashed where partly obscured.  White dashed line,
    tropopause. Modified from \citet{WoKi94}}
  \label{f.example_redoubt}
\end{figure}

\begin{figure}[!htb]
  \centering
  \includegraphics[width=0.8\textwidth]{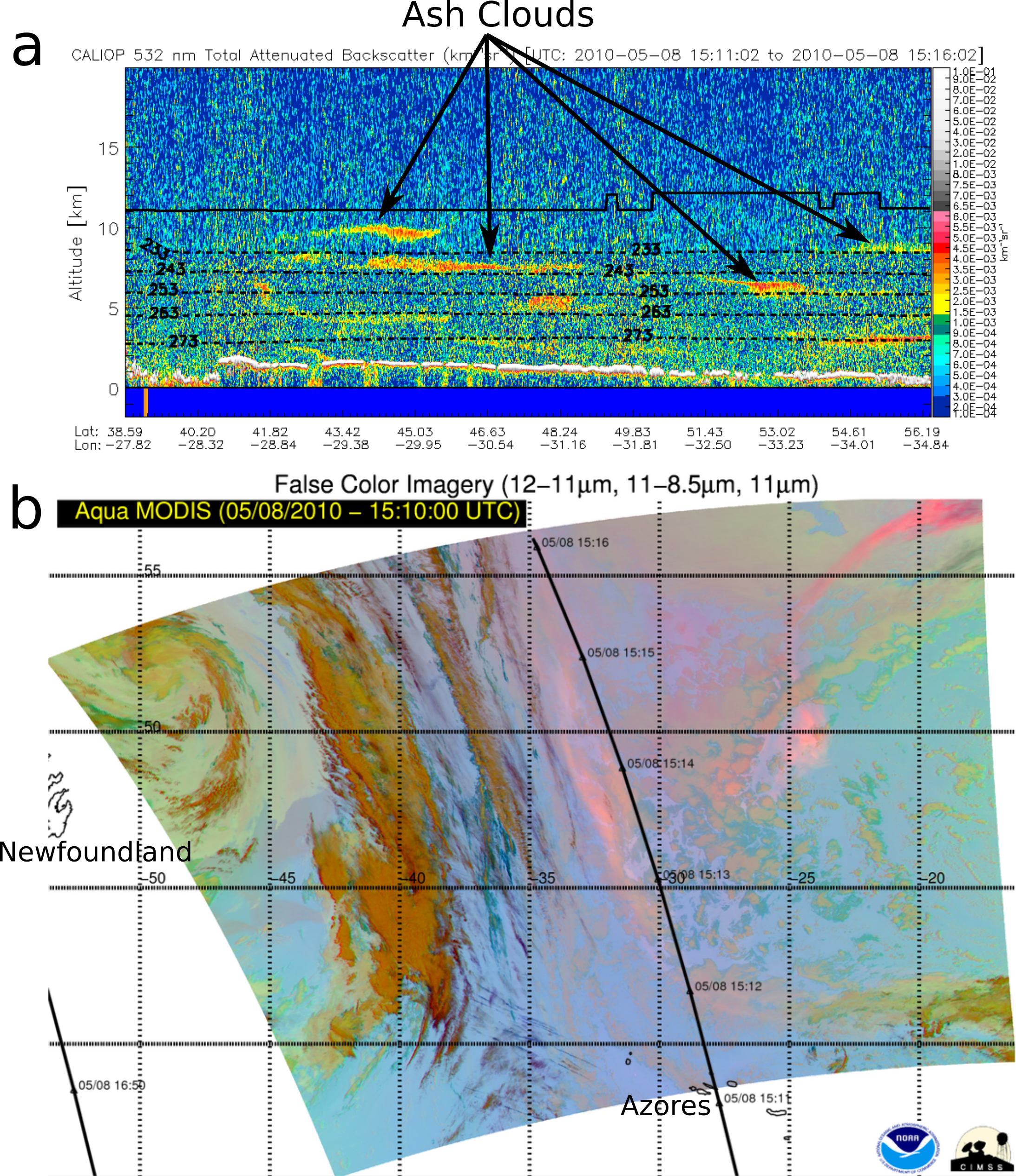}
  \caption{\textbf{(a)} CALIOP nadir LIDAR total attenuated backscatter (along
    track shown in (b)) showing complex layering of Eyjafjallajokull
    ash cloud on 8 May, 2010, isotherms (kelvins, black dotted lines)
    and tropopause (black solid line). \textbf{(b)} Aqua-MODIS RGB false color
    image \citep{doi:10.1002/jgrd.50173} of North Atlantic
    capturing this ash cloud (pink hues).}\label{f.example_eyja}
\end{figure}

Photography and satellite imagery of numerous volcanic eruptions show
that stratification or layering of volcanic clouds is a fundamental
aspect of volcanic cloud development (Fig. \ref{f.example_redoubt}).
Lidar backscatter data have been key in defining this layered
structure in distal regions (Fig. \ref{f.example_eyja}).  Volcanic
layers can be stratospheric as well as tropospheric.

  The layers form and separate by numerous processes, some unique to
  volcanic clouds.  The primary volcanic cloud layer near the vent,
  such as the volcanic umbrella cloud or anvil cloud, arises from the
  driving of hot eruptive gas and ash parcels outward around their
  equilibrium level, or neutral buoyancy height \citep{SpBuCa97}.  Ash
  accretion, ash re-entrainment, source variability -- injection of
  ash at different altitudes with changing eruption rate and wind
  fields, and gas-ash separation cause development of multiple layers
  \citep{HoSeWo96, tupper2004evaluation, thorsteinsson2012high}.
  Double diffusion and convective sediment flux to a single
  \citep{WoKi94, Bu98, HoBuAt99a, HoBuAt99b, CaJe12} and multiple
  \citep{CaJe13} levels by descending fingers that intrude below the
  level of a major volcanic cloud layer have been observed, and
  recreated in the laboratory.  As suggested by brightness
  temperatures over the surface of near-vent clouds and ground based
  photography, umbrella clouds can be solitary, accompanied by a
  single lower intrusion resulting from re-entrainment and column-edge
  downflow \citep{WoKi94, Bu98} or accompanied by lower level skirt
  clouds \citep{barr1982skirt}, which may or may not contain ash.
  Mechanical unmixing of particulate-laden and gaseous volcanic cloud
  components has been noted as a further cause of volcanic layer
  formation \citep{hws96, fero2009simulating}, perhaps enhanced by
  gravity current slumping of the particle laden component
  \citep{prata2017atmospheric}.

  In contrast to the well-defined and complex layers discussed above,
  volcanic ash transport and dispersal models (VATDs), used to
  forecast ash cloud motion, display layering of only a single ash
  cloud, if derived from wind shear; or after the fact from a
  numerical inversion for variable source height, which is otherwise
  assumed a fixed parameter
  (\citep{kristiansen2015stratospheric}. Although horizontal, planview
  resolution can be good, VATDs thus have difficulty in reproducing
  the vertical thickness and multilayering of distal volcanic clouds
  \citep{DEVENISH2012152, FOLCH2012165, HEINOLD2012195}.  In VATDs,
  dispersal in the vertical direction is commonly described by a
  single vertical diffusivity, $\kappa_z$, which in many models is
  taken to have the same value as the horizontal diffusivity,
  $\kappa_h$; the result is then uniform or isotropic ash dispersion.

\hypertarget{problem-statement}{%
\subsection{Problem Statement}\label{problem-statement}}

The formation mechanism and morphology of distal ash cloud layers are
poorly understood, and one potential mechanism for their formation and
characteristics is the subject of the present contribution. Our
working hypothesis is that, with particle settling, the structure of
the atmosphere itself can cause layer formation, through the process
of enhanced suspension in vertically restricted regions of high
turbulence. Given that the mechanism explored herein results from the
properties of the atmosphere itself, it may be the primary, distal
layer-forming mechanism. Understanding the reason for the occurrence
and morphology of specific ash-rich cloud layers is critical for
correctly characterizing extended persistence of ash in the
atmosphere, especially when multiple separate layers occur
together. Satellite sensors penetrate only partially into the highest
layer of an optically thick cloud, and satellite remote sensing
algorithms are most sensitive to the column integrated ash properties,
when multiple optically thin layers are present. A correct
understanding of layer formation and morphology is also critical for
any attempts to construct VATD models capable of producing output
consistent with observations of vertical dispersion.

\section{Materials and Methods}




\hypertarget{data}{%
\subsection{Background Data}\label{data}}

\hypertarget{atmosphere}{%
\subsubsection{Atmosphere}\label{atmosphere}}

Volcanic clouds both create turbulence and are subject to ambient
atmospheric turbulence.  In volcanic clouds near the vent, turbulence
is created by both the Rayleigh-Taylor and Kelvin-Helmholtz
mechanisms, as the clouds intrude into the atmosphere as gravity
currents.  Kelvin-Helmholtz instability is driven by the shear between
the intruding cloud and the atmosphere \citep{britter_simpson_1981}.
Rayleigh- Taylor instability is driven by convective sedimentation,
fingering and local, eddy scale density reversal \citep{WoKi94, hws96,
  chakraborty2006volcan}.

Information of sufficient resolution in the vertical direction to
discover and characterize the layered structure of the atmosphere is
obtained from airborne measurement campaigns or rawinsonde balloon
releases \citep{dehghan2014comparisons, cho2003characterizations,
  pavelin2002airborne}. Several methods have been developed to derive
turbulence from rawinsonde and other high-resolution data.
\citet{VaVa98} measured changes in the refractive index structure
parameter for radio waves, as turbulence causes changes in the
refractive index, based on rawinsonde pressure, temperature, humidity,
wind speed and wind direction data. \citet{clayson2008turbulence} used
variations in the potential temperature profile from an idealized
profile to calculate the Thorpe scale, and derive turbulent
dissipation and diffusivity.  These estimates can be made for single
rawinsonde profiles with simple calculations. There are drawbacks of
course to extrapolating such high-resolution or point data to a
regional scale because of spatial and temporal inhomogeneity; the
troposphere is highly transient and spatially variable
\citep{clayson2008turbulence, thouret2000general}.  Tropospheric
isobaric surfaces are not necessarily parallel to the earth's surface,
especially at fronts and in mountain waves \cite{ShTrLa12}. Fronts are
associated with tropopause folds, non-horizontal isobaric surfaces
separating cold from warm air, and turbulence in folds is generated by
local dynamic and convective instabilities. Mountain waves form as the
density stratified atmosphere flows past the lee side of a mountain or
mountain range.  These waves can break, resulting in local turbulence
concentrated in non-horizontal layers.

  In the free atmosphere, parameters such as moisture content and
  temperature do not change monotonically with height; there are
  regions of relatively homogeneous, convecting or turbulent air,
  separated by regions in which parameters vary rapidly \citep{VaVa98,
    ShTrLa12}.  Both the stratosphere and the troposphere are layered
  on scales of $\mathcal{O}[0.1-1km]$ \citep{MaFuYa93, WiLuHa14}, but
  the layers in the troposphere tend to be more transient and
  discontinuous \citep{gage1980use}.  In the troposphere, layers of
  high turbulence, including high vertical turbulence, can be
  indicated by constant relative humidity (RH) or mixing ratio, $q$
  \citep{cho2003characterizations}.  RH is thus an important proxy for
  turbulence intensity.  (Atmospheric moisture is also important in
  aiding plume lift, especially in plumes from weak sources, or at low
  latitude, where the moisture content is high \citep{SpBuCa97,
    TuTeHe09}).  As a result of the layering, large-volume or bulk
  turbulence is highly anisotropic ($\kappa_h \gg \kappa_z$), and only
  within thin, well-defined layers is it approximately isotropic
  ($\kappa_{h, local} \approx \kappa_{z, local}$) \citep{gage1980use}.

\hypertarget{ash-clouds}{%
\subsubsection{Ash Clouds}\label{ash-clouds}}

Data are available from a number of sources on the shape and structure
of both near-vent and distal ash clouds.  In the near-vent region,
data tend to be more limited, due to the greater optical depth.
Nevertheless, cloud brightness temperature (BT) as measured from
nadir-looking geostationary and low-earth orbiting satellites provides
much useful information, as the topography of the top of the near-vent
clouds can be quite variable, with a distinct high point or swell
above the central vent that might be many kilometers above the top of
the main umbrella cloud \citep{fero2009simulating}.  In addition,
airborne and ground-based photography and videography have provided
extensive data on the features at the base of the main umbrella or
anvil, and within the underlying cloud layers.  Visible satellite
imaging of the near vent cloud top consistently reveals strong,
well-defined three-dimensional vortex structures above the vent, which
evolve to smooth, somewhat more diffuse structures in the umbrella
cloud \citep{PoBuJo16}.

\begin{figure}[!htb]
  \centering
  \includegraphics[width=0.8\textwidth]{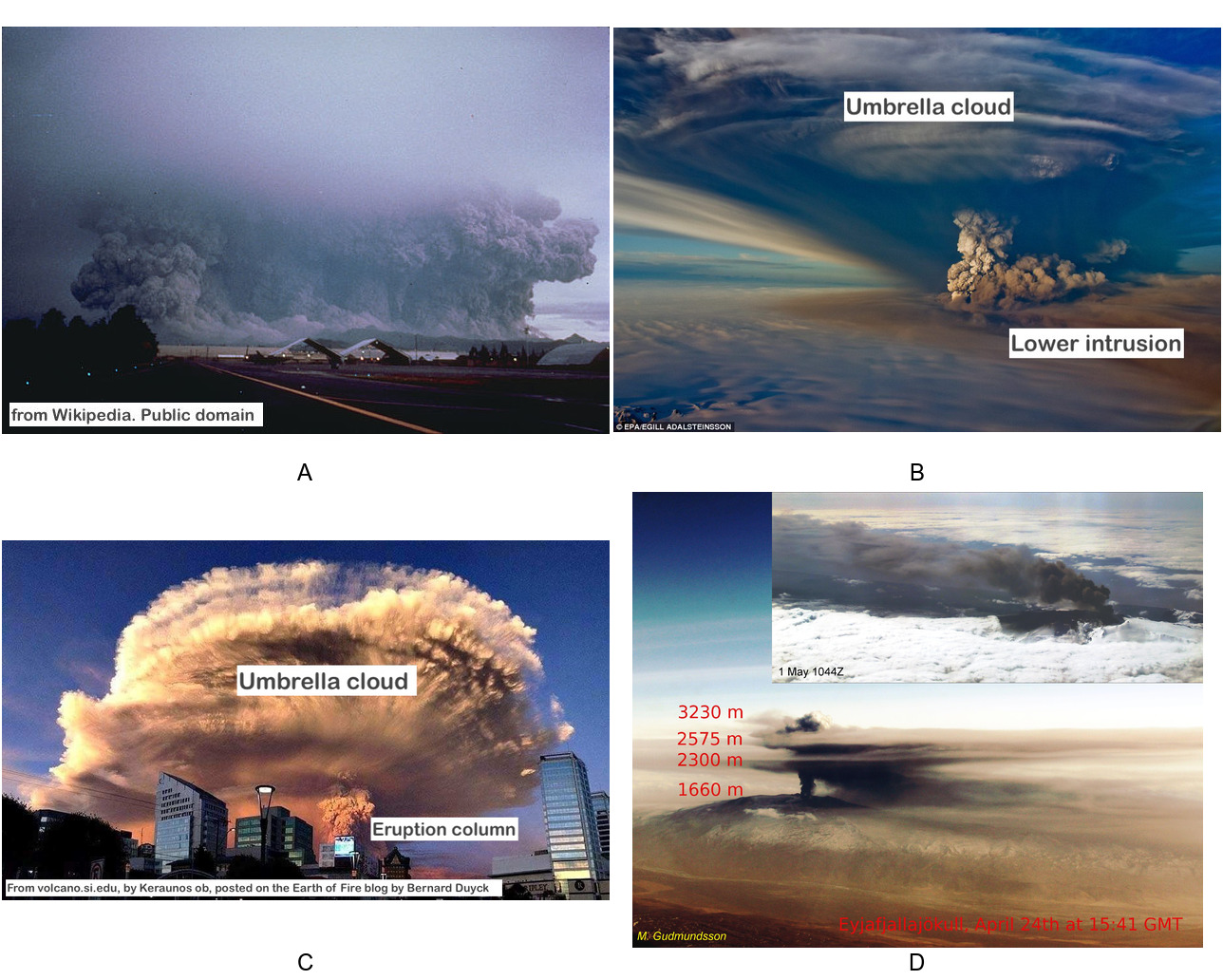}
  \caption{\textbf{(a)} Eruption of Pinatubo, June 15, 1991.  Emission
    from vent as well as pyroclastic flows results in ash injecting
    into the atmosphere at all heights up to 40 km. In the public
    domain. \textbf{(b)} Eruption of Grimsvotn, May 22, 2011. Ash
    injected in umbrella cloud (above), and in lower intrusion from
    secondary, ash-rich cloud (below). Umbrella at 18-20 km.  Modified
    from
    \url{https://www.dailymail.co.uk/news/article-1389846/Iceland-volcano-eruption-2011-Grimsvotn-hurles-ash-plume-12-miles-sky.html}. \textbf{(c)}
    Eruption of Calbuco, April 22, 2015.  Ash injected in single
    umbrella cloud at c. 15 km. Modified from
    \url{https://volcano.si.edu/volcano.cfm?vn=358020}.  \textbf{(d)}
    Eruption of \eyja, April 24 and May 1, 2010. Ash injected at
    multiple heights in relatively quiescent atmosphere on April 24.
    On May 1, under windy conditions, ash mostly injected from single
    downwind plume. Modified from photographs by M. Gudmundsson.
  }
  \label{f.near_vent}
  \end{figure}

Although the air can be choked with opaque, diffuse ash bodies that
extend to ground level near vent (Fig. \ref{f.near_vent}a, b), and
although gravitational intrusions, such as umbrella clouds, are
wedge-shaped by nature and therefore of variable depth
(Fig. \ref{f.near_vent}c), measurements have been made of the
$\Delta$BT between cloud top and edge of the main cloud.  The results
suggest that umbrella clouds at the vent typically encompass depths of
$\mathcal{O}[5]$ km (Fig. \ref{f.example_redoubt}; Table
\ref{t.proximal_depth_ranges}), which makes a large mass of ash available for
transport at sometimes high levels.  Some of the lower near-vent
clouds also become distal clouds (Fig. \ref{f.near_vent}d).  In some
cases, therefore, it might be possible to find the entire troposphere
and even lower stratosphere charged with ash, or only a distinct layer
or two of $\mathcal{O}[5]$ km depth.

\begin{table}[!htb]
  \caption{Measured main, near-vent, upper volcanic cloud depth from top to cloud
    edge.  Measured from geostationary imagery in first scene after
    eruption start (photography for Redoubt).  Data from
    \citet{bear2020automated}; and \citet{pouget2013estimationa} and
    \citet{HoSeWo96} (below divider)} \label{t.proximal_depth_ranges}
  \centering
\begin{tabular}{p{1in}p{2in}p{1in}p{1in}p{1in}}
  \hline
  \textbf{Volcano}	& \textbf{Eruption start date}	&
                                                         \textbf{Height,
                                                         km ASL} &
                                                                   \textbf{Depth,
                                                                   km}
  & \textbf{No. layers} \\
  \hline
  Tinakula	& Oct 20, 2017, 2350 UT			& 16.6 & 4.9 &
  1 \\
  Tinakula	& Oct 20, 2017, 1930 UT			& 15.1 & 3.4 &
  1 \\
  Rinjani	& Aug 1, 2016, 0345 UT	        	& 5.5	& 4.0
  & 1\\
  Manam		& Jul 31, 2015, 0132 UT			& 13.7 &
                                                                 9.2$\dagger$
  & 2 \\
  Sangeang Api	& May 11, 2014, 0832 UT			& 15.4 &
                                                                 12.0$\dagger$
  & 2 \\
  Kelud		& Feb 13, 2014, 1632 UT			& 15.3 & 2.8 &
  1\\
  Manam		& Jan 27, 2005, 1400 UT			& 24.0 & 3.0 &
  1 \\
  Manam		& Oct 24, 2004, 2325 UT			& 18.5 & 1.5
                                                                 & 1\\
  \hline
  Pinatubo & Jun 15, 1991, 2241 UT & 23.6 & 4.7 & 1\\
  Redoubt & Apr 21, 1990, 1412 UT & 12.0 & 4.9 & 2 \\
  Mount St. Helens & May 18, 1980, 2020 UT & 13.0 & 1.0 & 1 \\
  \hline
\end{tabular}
  $\dagger$ Depth possibly overestimated.
\end{table}

In the distal region, airborne lidar, EARLINET-AERONET and CALIOP data
typically show much thinner, more discontinuous cloud structures
(Fig. \ref{f.example_eyja}).  Three separate tabulations of distal ash
cloud layer data for the \eyja ~plume have been published
\citep{jwbdfc96, AnTeSe10, winker2012caliop}. These data suggest that
distal clouds from this tropospheric eruption were typically 0.3-3 km
thick, made up of 2-3 layers, with individual layers of 0.3-1.4 km
depth, and maximum age of 129 h (< 1 week) (Table
\ref{t.depth_ranges}). \citet{vernier2013advanced} using CALIOP data,
discerned two or more well-defined layers in the cloud from
Puyehue-Cordon Caulle three weeks after the eruption. Some of the
layers showed fold or wrap-around structures (Figure 1a in
\citet{vernier2013advanced}), perhaps related to vertical-plane
chaotic mixing \citep{pierce1993chaotic}.  Clouds were up to 3 km
thick, with individual layers of 0.1-2 km depth, in the upper
troposphere-lower stratosphere (UTLS), centered on the tropopause at
8-14 km altitude.  On July 12-13, 1991, 26 days after the last major
eruption of Pinatubo, a lidar flight noted numerous stratospheric
layers \citep{winker1992preliminary}.  The data showed a number of
well-defined layers of 0.5-1 km depth between about 14 and 25 km
altitude (Figure 1 in \citet{winker1992preliminary}).  Along much of
the line of flight there were two layers, but in places there were up
to five.  In all studies cited above, distal ash layers were
horizontal or tilted relative to the horizon, and had extents in the
cross-transport direction of hundreds of km in the troposphere
(\eyja), to thousands of km in the stratosphere (Puyehue, Pinatubo).
Ash is retained longer in the stratosphere than in the troposphere as
suggested by the data cited herein.

  Once the particles have propagated far from vent (>11 hr in the case
  of a large eruption, e.g., Pinatubo), they no longer retain
  significant memory of source conditions \citep{fero2009simulating}.
  Back trajectories of distal ash clouds for Eyjafjallajökull and
  Puyehue-Cordon Caull\'e are generally consistent with theoretically
  possible cloud heights at the source \citep{winker2012caliop,
    vernier2013advanced}.  However, it is not clear that ash was
  injected at these altitudes at the source, given uncertainties in
  vertical parcel motion and settling speed, or lack of incorporation
  thereof in the models \citep{MaPoSi14, vernier2013advanced}. It is
  at least possible that some cloud layers were generated at distance
  from the volcanic source.

\begin{table}[!htb]
  \caption{Measured distal \eyja~ cloud depths from CALIOP
    lidar.  Data from \citet{winker2012caliop} (top),
    \citet{marenco2011airborne} (middle) and
    \citet{schumann2011airborne} (lower section). Note that number of
    layers varies with spatial position.} \label{t.depth_ranges}
  \centering
\begin{tabular}{p{1in}p{1in}p{1in}p{0.8in}p{0.8in}p{0.8in}}
  \hline
  \textbf{Date} & \textbf{Cloud}	& \textbf{Height range}	& \textbf{
                                                                          Depth} &
                                                                                       \textbf{Age, hr}
  & \textbf{No. layers} \\
  & & \textbf{km ASL} & \textbf{km} & & \\

  \hline
  Apr 15 & 20100415 & $1.41-3.23$ & 0.51 & $<6$ & -- \\
  Apr 16 & 20100416-a & $3.77-5.50$ & 0.58 & 30 & $>1$\\
  Apr 16 & 20100416-b & $1.97-7.27$ & 0.67 & 24 & $>1$\\ 
  Apr 17 & 20100417-a & $0.20-6.28$ & 0.76 & 42 & 1 \\
  Apr 17 & 20100417-b & $0.05-4.00$ & 0.61 & 42 & 1 \\
  Apr 18 & 20100418-a & $3.14-5.59$ & 0.81 & 66 & -- \\
  Apr 18 & 20100418-b & $3.75-6.49$ & 0.86 & 66 & -- \\
  Apr 19 & 20100419-a & $3.20-5.26$ & 1.06 & 71 & -- \\ 
  Apr 19 & 20100419-c & $2.48-3.94$ & 0.45 & 30 & -- \\ 
  Apr 19 & 20100419-d & $4.63-5.20$ & 0.41 & $114–126$ & -- \\
  Apr 20 & 20100420 & $0.05-1.88$ & 1.08 & $20–24$ & -- \\ 
  \hline
  May 4 & 20100504 & $2.3-5.5$ & 0.5 & -- & $1-2$ \\
  May 5 & 20100505 & $2.4-4.5$ & 0.9 & -- & $1-2$ \\
  May 14 & 20100514 & $5.1-8.1$ & 1.1 & -- & $1-3$ \\
  May 16 & 20100516 & $3.4-5.5$ & 1.2 & -- & $1-3$ \\
  May 17 & 20100517 & $3.5-5.6$ & 1.3 & -- & $1-3$ \\
  May 18 & 20100518 & $2.5-4.9$ & 0.9 & -- & $1-3$ \\
  \hline
  Apr 19 & 20100419-1 & $3.9-5.6$ & 1.7 & $105-111$ & $>1$ \\
  Apr 19 & 20100419-2 & $3.5-3.8$ & 0.3 & $104-108$ & 1 \\
  Apr 19 & 20100419-3 & $3.9-4.2$ & 0.3 & $105-108$ & 1 \\
  Apr 22 & 20100422-4 & $0.7-5.5$ & -- & $49-50$ & diffuse \\
  Apr 23 & 20100423-5 & $2.1-3.4$ & 1.3 & $40-58$ & $>1$ \\
  May 2 & 20100502-6 & $1.6-3.7$ & 2.1 & $7.1-12$ & $>1$ \\
  May 9 & 20100509-7 & $3.5-4.9$ & 1.4 & $97-129$ & 1 \\
  May 13 & 20100513-8 & $2.8-5.4$ & $0.4-0.7$ & $71-78$ & 1 tilted \\
  May 16 & 20100516-9 & $3.6-7.0$ & 3.4 & $58-66$ & $>1$ \\
  May 17 & 20100517-10 & $3.2-6.3$ & 3.1 & $66-88$ & $>1$ \\
  May 18 & 20100518-11 & $2.8-3.4$ & 0.6 & $81-100$ & 1 \\
  May 18 & 20100518-12 & $4.0-5.7$ & 1.7 & $66-78$ & $>1$ \\
  \hline
\end{tabular}
\end{table}

The data presented herein, in text, Figures \ref{f.example_eyja},
\ref{f.near_vent}, and Tables \ref{t.proximal_depth_ranges},
\ref{t.depth_ranges}, suggest that the number of layers increases with
time, and the depth decreases (see also \citet{dacre2015volcanic}).
Volcanic source conditions are eventually lost after an eruption,
meaning that the atmosphere completely controls cloud shape. Layers
are more transient in the troposphere than in the stratosphere, as
mentioned for certain eruptions (also, \textit{cf.}  data in Table
\ref{t.proximal_depth_ranges} with \ref{t.depth_ranges}; \citep{
  thouret2000general}).  In summary, the features of volcanic clouds
include the following:
\begin{enumerate}
  \def\labelenumi{\arabic{enumi}.}
\item Near-vent clouds often evolve from sharply defined with clear
  eddy or vortex structure to diffuse or diaphanous, into distal, thin
  well-defined layers with sharp, smooth boundaries
\item There are often multiple distal layers
\item Collocation in position in planview is common (stacking), but
  not pervasive
\item Sometimes distal cloud forms are horizontal; sometimes sloping
  or tilted relative to horizontal
\item Separate particle clouds exist, not only separate gas and ash
  clouds
\item Horizontal extent $\gg$ vertical extent
\item Vertical extent of near-vent layers is $\mathcal{O}[5]$ km,
  which with time results in
\item Vertical extent of single, distal layers being
  $\mathcal{O}[0.1 - 1]$ km
\end{enumerate}
The present contribution seeks to provide explanations for some of
these features.

\hypertarget{model}{%
\subsection{Model}\label{model}}

The advection-diffusion equation forms the basis for all VATD
models. To illustrate differences and potential problems in
implementation of VATD and the underlying physics, in this section, we
introduce two zeroth-order simplifications.  Our goal is to contrast
the basic behavior of an ash cloud under conditions of isotropic
(or constant $\kappa_h$ and $\kappa_z$) turbulence, as assumed in the
models, and layered turbulence, as we find in the atmosphere.  We base
our modeling on generation of synthetic atmospheres with and without
multiple turbulent layers separated by relatively quiescent air.

In the case of isotropic turbulence, we begin by assuming Cartesian
coordinates, $(x, y, z)$, with velocity components, $(u, v, w)$.  The
three components of the turbulent diffusivity,
$(\kappa_x, \kappa_y, \kappa_z)$ are the same, $\kappa$.  The
concentration of particles in the $i$-size fraction, $C_i$, varies
in time, $t$ and space as:
\begin{equation}
  \frac{\partial C_{i}}{\partial t}+\frac{\partial}{\partial x}\left(u
    C_{i}\right)+\frac{\partial}{\partial y}\left(v
    C_{i}\right)+\frac{\partial}{\partial z}\left(w C_{i}\right) =
  \frac{\partial^{2}}{\partial x^{2}}\left(\kappa C_{i}\right) +
  \frac{\partial^{2}}{\partial y^{2}}\left(\kappa C_{i}\right) + \frac{\partial^{2}}{\partial z^{2}}\left(\kappa C_{i}\right)+\Phi
\end{equation}
where $\Phi$ represents the source/sink function, which in the case of
ash clouds is mostly represented by aggregation and disaggregation of
small particles.  In the present case, such processes are set to zero.
We assume a two-dimensional system with a point-source in time and
space, $w=w_s$, the settling speed, and that, following a streamtube,
the motion of the volcanic cloud can be characterized by a single
downwind coordinate direction $s$ -- for which the axis is everywhere
tangent to the plume centerline, e.g., \citep{w77,hb85} -- and speed
$U$ in that direction. Under these assumptions, the
advection-diffusion equation becomes:
\begin{equation}
  \frac{\partial C_i}{\partial t} + \frac{\partial}{\partial s}\left(U
    C_{i}\right)+\frac{\partial}{\partial z}\left(w_{s} C_{i}\right) = \frac{\partial^{2}}{\partial s^{2}}\left(\kappa C_{i}\right) + \frac{\partial^{2}}{\partial z^{2}}\left(\kappa C_{i}\right)
\end{equation}
with the well-known solution for the impulse initial condition
\citep{Cs80, RoWe02turbulent}:
\begin{equation}
  C_i(s, z, t) = \frac{C_{i0}}{4 \pi \kappa t}\exp\left[-\frac{(s - s_0 - Ut)^2
      +(z - z_0 - w_st)^2}{4 \kappa t}\right]. \label{e.greenfunction}
    \end{equation}
It is reasonably clear that the solution is a Gaussian in $(s, z)$,
in which ash spreads, settles and is blown downwind with time.

In the second case, of layered turbulence, more realistic for the free
atmosphere, we assume particles filling a layer of finite vertical
extent.  Due to its internal turbulence, concentration varies in $t$
but not in $s$ or $z$ within the layer, $C_i(s, z, t) = C_i(t)$.
There is no flux at the upper boundary of such a layer, as any
particles injected upwards by eddies will settle back down into the
layer.  There is a flux boundary condition at the lower boundary where
$\kappa_z \rightarrow 0$, and in this case, the advection-diffusion
equation becomes, at the lower boundary:
\begin{equation}
\frac{\partial}{\partial t}\left(
  C_{i}\right)+\frac{\partial}{\partial z}\left(w_{s} C_{i}\right) = 0 \label{e.boundary}
\end{equation}
If furthermore it can be assumed that some particles near the cloud
base fall from the turbulent layer when their weight overcomes the
internal turbulence at the lower cloud edge, thus developing a
step-like concentration gradient at the base of the streamtube, then:
\begin{equation}
C_{i}(t, z)=H(z) C_{i}(t)
\end{equation}
where $H(z)$ is the Heaviside step function, then:
\begin{equation}
\frac{\partial}{\partial t}\left(
  C_{i}\right)+\frac{\partial}{\partial z}\left(w_{s} C_{i}\right) = 
\frac{\partial C_{i}}{\partial t}+w_{s} C_{i} \frac{\partial {H}(z)}{\partial z}=0
\end{equation}

Integrating through the layer depth, $h$:
\begin{equation}
 \frac{\partial C_{i}}{\partial t} \int_{0}^{h} dz = -w_{s}C_{i} \int_{0}^{h} \delta(z) dz
\end{equation}
we obtain:
\begin{equation}
\frac{d C_{i}}{d t}=-\frac{w_{s}}{ h} C_{i}
\end{equation}
which has solution:
\begin{equation}
  C_i = C_{i0} \exp\left(- \frac{w_s (t-t_0) }{ h}\right)
  \label{e.hazen}
\end{equation}

In the quiescent layer below the boundary, particles only settle and
are advected downwind, there is no turbulence mechanism to enhance
persistence within the layer.  Thus, turbulent layers can retain
particles longer than do quiescent layers because of continuing
re-entrainment in eddies. Particles fall relatively rapidly through
the quiescent layers because of uninhibited settling, sometimes even
enhanced by the effects of convective sedimentation \citep{HoBuAt99a},
which is not included in the present model.

Based on similarity theory, the timescales for the processes under the
different particle transport conditions arising in different layers
can be used to examine the conditions under which diffusion or
settling dominates.  From Eq \ref{e.greenfunction}, the timescale of
vertical diffusion, $\tau_1$, through a layer of depth, $h$, is
given as $ \tau_1 = \frac{h^2}{\kappa}$.  From Eq \ref{e.hazen}, the
timescale of settling through the same layer, $\tau_2$, is
$ \tau_2 = \frac{h}{w_s}$.  The ratio of the two timescales indicates
domination of particle transport by settling or dispersion in the
vertical direction.  The ratio is given by the dimensionless
group,$\Pi_1$:
\begin{equation}
  \Pi_1 = \frac{\tau_1}{\tau_2} = \frac{h w_s}{\kappa} \label{e.pi}
\end{equation}

In the following section, we explore results from these
simplifications and the similarity analysis, as well as numerical
solutions to more complicated cases.  Numerical solutions are provided
for the Ash3D VATD model \citep{schwaiger2012ash3d}, as well as both
Eulerian and Lagrangian model codes (Table \ref{t.experiments}).

\section{Results}

\begin{figure}[!htb]
  \centering
  \includegraphics[width=0.9\textwidth]{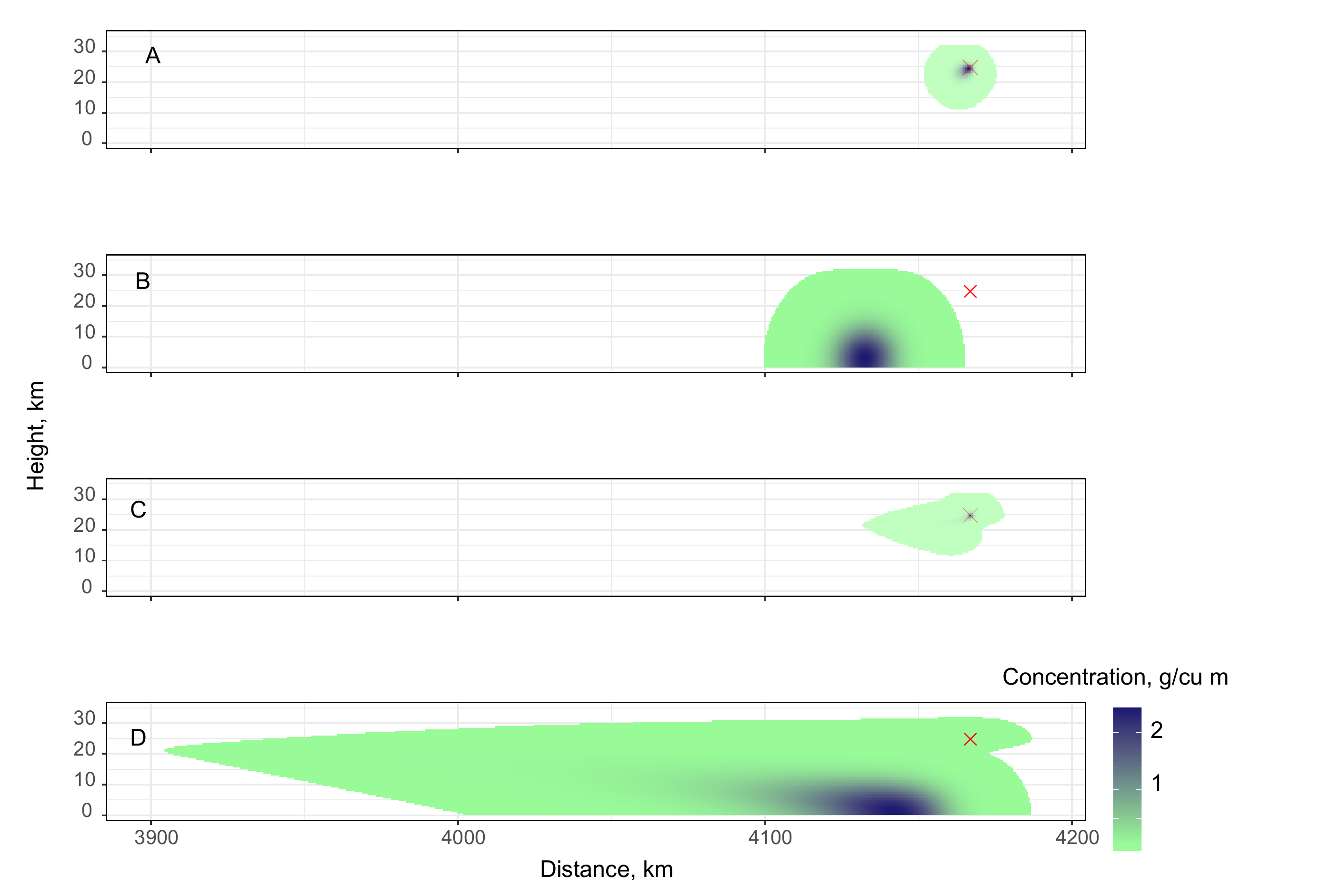}
  \caption{Cross sections through Ash3D output, with: \textbf{(a)}
    instantaneous source, showing advection, settling and isotropic
    dispersal of ash. 10 min after release. \textbf{(b)} instantaneous
    source, showing advection, settling and isotropic dispersal of
    ash. 120 min after release. \textbf{(c)} maintained source,
    showing advection with wind shear, settling under isotropic
    turbulence. 10 min after release. \textbf{(d)} maintained source,
    showing advection with wind shear, settling in isotropic
    turbulence. 120 min after release. Note that maintained source and
    wind shear result together in elongated dispersal pattern, and
    layer development. Red cross, source
    location.}\label{f.ash3d_xsxn}
\end{figure}

Following from Eq \ref{e.greenfunction}, spread from a point source in
a VATD model, with isotropic turbulence and a wind of constant speed
with height, is shown in Figure \ref{f.ash3d_xsxn}a, b.  Ash diffuses
and progressively spreads from the source as the center of mass
descends at the settling speed.  Using a higher settling speed, the
rate at which the center of mass descends increases, but the rate at
which the particles disperse from the center of mass remains constant.
Thus, at any one height below the source, particles with a higher
settling speed should be spread less distance from the source.

\begin{landscape}
\begin{table}[!htb]
  \caption{Simulation parameters. Duration refers to emission from
    vent.  } \label{t.experiments}
  \centering
  \begin{tabular}{p{1.5in}p{1.5in}p{1.5in}p{1.5in}p{1in}p{1in}}
    \textbf{Parameter} & \textbf{$\kappa=$const} &
                                                               \textbf{$\kappa=$const}
    & 
        \textbf{Layered Atmosphere} &  & \\
    & \textbf{wind=const} & \textbf{windshear} & & & \\
    \hline
    Simulation type & Ash3D & Ash3D & Lagrangian-1 & Lagrangian-2 & Eulerian \\
    Source type & Point & Point & Point & Point & Layer \\
    Source height, km & 24.75 & 24.75 & 24.75 & 4.2 & $3.8-4.2$ \\
    Particle size, $\mu$m & 1000 & 1000 & 1000 & 40 & 40 \\
    Settling speed, m/s & 3 & 3 & 3 & 0.02 & 0.02 \\
    Amount & 0.01 km$^3$ & 0.01 km$^3$ & 1000 parcels & 1000 parcels & 10000 particles \\
    Duration & 0.2 hr & 0.2 hr & Instantaneous & Instantaneous & Instantaneous\\
    \hline
    Turbulent layer heights & -- & -- & $24-25$, $14.5-19$ km & $3.8-4.2$,
                                                            $2-2.7$ km &
                                                                       $3.8-4.2$,
                                                                       $2-2.7$
                                                                       km \\
    Wind speed & 5 m/s all elev & Below 19.75 km: 0 m/s & 5 m/s &
                                                                       10
                                                                       m/s &
                                                                             0 m/s \\
                       & & 19.75 to 21.75 km: 10 increasing to 50 m/s
    & & & \\
                       & & 21.75 to 23.75 km: 50 decreasing to 10 m/s
    & & & \\
                       & & Above 23.75 km: 10 m/s & & & \\
    \hline
  \end{tabular}
\end{table}
\end{landscape}
    
Dispersion in the presence of wind shear is shown in Figure
\ref{f.ash3d_xsxn}c, d. The shear distorts the dispersal pattern from
the idealized, spherically symmetric pattern seen in the constant wind
field, causing an elongation in the dispersal pattern centered at the
wind speed maximum. Because the deformation is by simple shear, cloud
thinning does not occur.  Wind-shear produced elongation thus creates
a volcanic cloud layer that continues to deepen by diffusion.

  In a layered atmosphere, we refer to Eq \ref{e.pi} to explore
  asymptotic behavior.  In layers for which $\Pi_1 > 1$, the diffusivity
  is low relative to the settling speed, the timescale of diffusion is
  therefore long, hence motion is controlled by settling.  In layers
  for which $\Pi_1 < 1$, the diffusivity is high relative to the
  settling speed, the timescale of settling is long, hence motion is
  dominated by diffusion.  Note also that as $h$ increases, the
  timescale of diffusion increases faster than does that for settling,
  meaning it becomes more likely the particles will exit a layer by
  settling than by diffusion.  For a typical diffusivity of
  $\kappa = 1$ m$^2$/s in the UTLS \citep{Wi04} at ~10 km altitude,
  and layer of depth $h = 1$ km, the critical settling speed,
  $w_{s, crit}$, dividing settling from diffusion dominated motion is
  c. 1 m/s, which would correspond to a pumice particle of diameter
  c. 100 $\mu$m at about 2400 kg/cu m (e.g., \citep{ScCoPr05}).

\begin{figure}[!htb]
    \centering
    \includegraphics[width=0.8\textwidth]{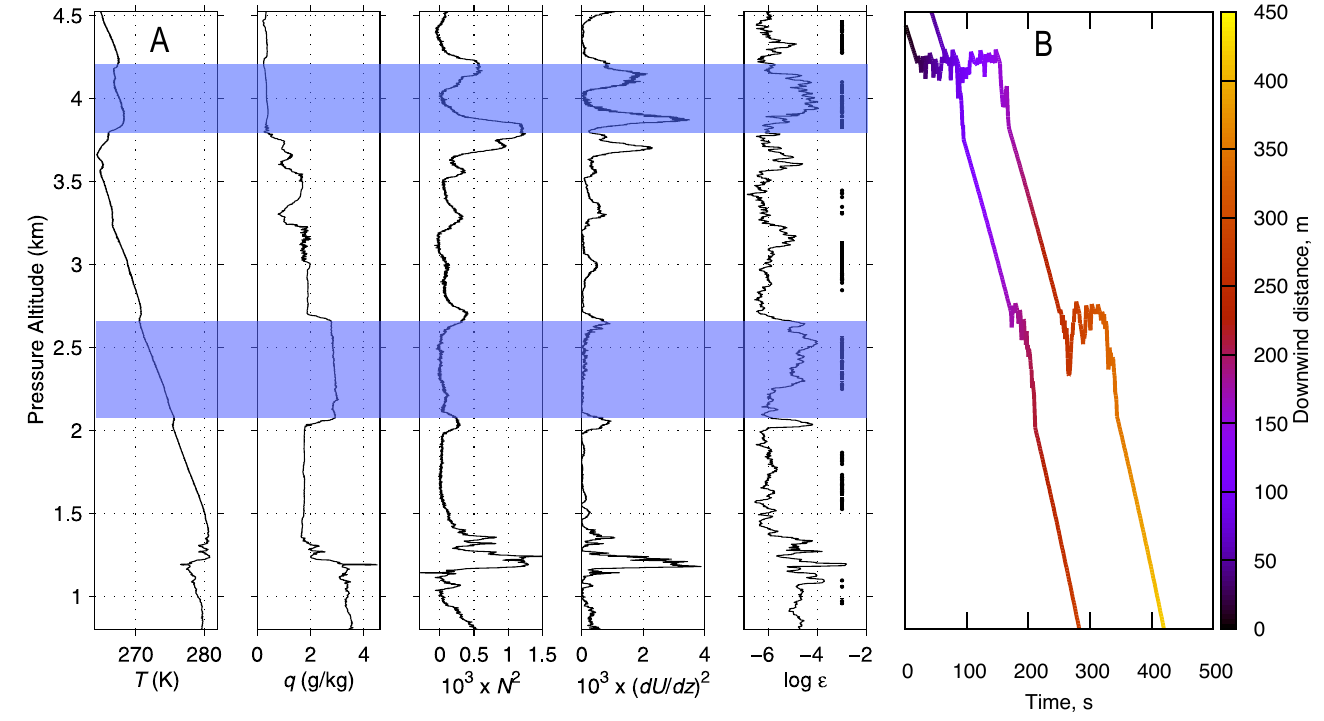}
    \caption{\textbf{(a)} Turbulent tropospheric layers (shaded)
      indicated by constant mixing ratio ($q$), high turbulent energy
      dissipation rate ($\log \varepsilon$), and bounded by high shear
      ($(dU/dz)^2$). Modified from
      \citet{cho2003characterizations}. Shaded layers are used in
      simplified layer models Lagrangian-2 and Eulerian (Table
      \ref{t.experiments}) \textbf{(b)} Lagrangian paths of
        two volcanic particles settling through the turbulent layers,
        in which suspension is enhanced. This is output from model
        Lagrangian-2.  \label{f.layers_Cho_2003}}
  \end{figure}

\begin{figure}[!htb]
\centering
  \includegraphics[width=0.7\textwidth]{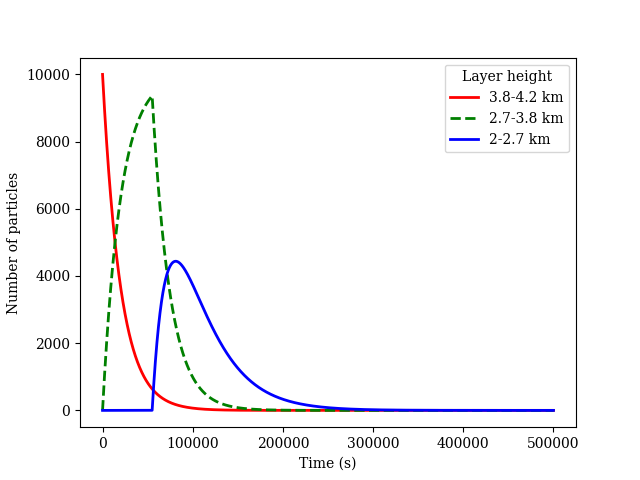}
  \caption{Eulerian model (Table \ref{t.experiments}) of
     number of particles in layered atmosphere.  Middle layer is not
     turbulent (green dashed curve); upper and lower layers are
     turbulent (red and blue solid curves) \label{f.settling}}
  \end{figure}

  Results for a simple layered system are shown in Figures
  \ref{f.layers_Cho_2003} and \ref{f.settling}.  These figures are
  simplified from the observations of
  \citet{cho2003characterizations}, who point out two layers of
  especially striking turbulence from 2-2.7 km and 3.8-4.2 km
  (Fig. \ref{f.layers_Cho_2003}a), but do not give specific values for
  turbulence intensity in any layers.  We therefore apply turbulence
  in these two layers through a Lagrangian random walk, and in other
  layers, no turbulence.  Particles in the turbulent layers, then,
  initiate the random walk, being ``stuck'' within the eddies of the
  turbulent layer (Fig. \ref{f.layers_Cho_2003}b).  Consider the lower
  boundary of a layer with strong turbulence.  All the particles there
  are subject to a random walk. They have a 50\% percent probability
  of going up, and a 50\% probability of going down.  Those particles
  sent above the lower boundary due to turbulence are sent to a
  position higher above the boundary than their original position at
  the boundary. This will give them a greater chance to spend a longer
  time in the turbulent layer, whether or not one considers
  settling. Thus, for those layers dominated by the random walk and
  dispersion, $\Pi_1 < 1$, Eq \ref{e.hazen} holds, and the behavior
  seen in Figures \ref{f.layers_Cho_2003} (purple layers) and
  \ref{f.settling} (red and blue lines) occurs. Particles accumulate
  in the lowermost turbulent layers (e.g., blue curve in Figure
  \ref{f.settling}), and once reaching a peak, settle out only slowly.
  Thus, the lower, turbulent layers are the most likely ones in which
  to observe particles on longer timescales.

  The two modes of behavior, isotropic diffusion and layered
  diffusion, contrast starkly, as seen in explicitly comparing VATD
  output with that from the Lagrangian-1 model (Table
  \ref{t.experiments}; Fig. \ref{f.model_xsxn}).  Isotropic diffusion
  alone cannot lead to generation of ash layers away from source, but
  layered diffusion can, even in a constant windfield.

\section{Discussion and Conclusions}



In the present work, we have presented data and models on near-vent
and distal volcanic cloud morphology and loading.  We have performed
numerical experiments comparing dispersal in an atmosphere with
constant $\kappa$ and wind, constant $\kappa$ and wind shear, and
variable $\kappa_z$ with height.  The observational data suggest that
more distal clouds of depth $\mathcal{O}$[0.1 to 1] km develop from
near-vent clouds of depth generally $1-5$ km.  The downwind clouds
occur at heights consistent with the original eruption column heights
for both tropospheric and stratospheric eruptions. The depth range of
the distal layers, being markedly less than the near-vent depth range,
and the common stacking of more distal ash cloud layers, suggest that
their development is controlled by atmospheric processes.  The
observations are consistent with the working hypothesis that the
layering of the atmosphere in turbulence intensity, causing
alternating suspension and settling dominated behavior of particles,
is a cause of distal layer morphology.

\begin{figure}[!htb]
\centering
  \includegraphics[width=0.9\textwidth]{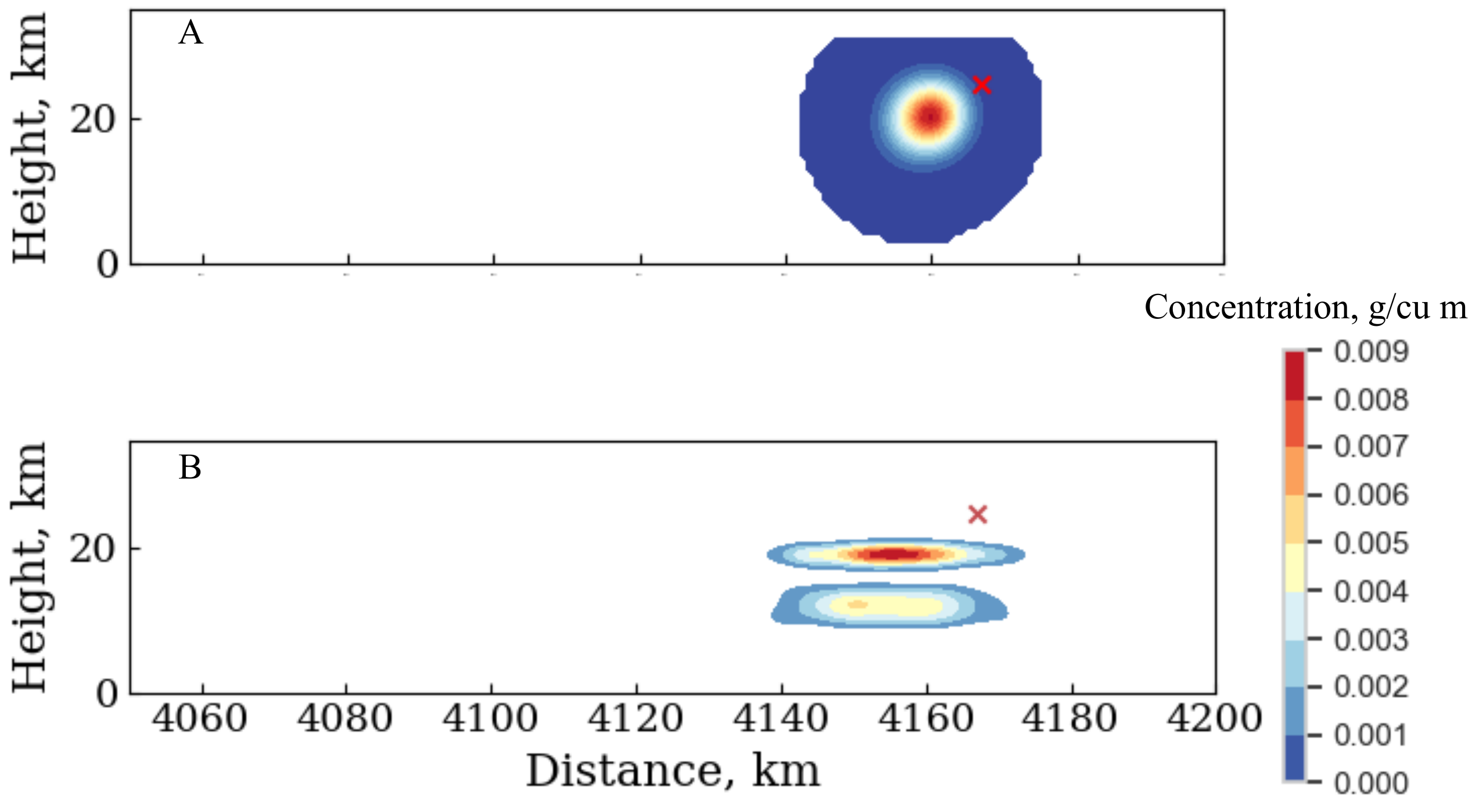}
  \caption{\textbf{(a)} Cross section through Ash3D output (Table
    \ref{t.experiments}), with instantaneous source and no wind shear,
    showing advection, settling and isotropic dispersal of ash.
    \textbf{(b)} Cross section through Lagrangian-1 dispersion model
    (Table \ref{t.experiments} with turbulence layered atmosphere,
    showing advection, settling and non-isotropic ash dispersal. Red
    $\times$ is point of origin for particles, and color gradient is
    scaled to concentration in both. }\label{f.model_xsxn}
\end{figure}

The present model outputs (Figs. \ref{f.ash3d_xsxn},
\ref{f.model_xsxn}) are consistent with those of VATD and
backtrajectory models that layers can be produced under conditions of
certain source characteristics or wind shear \citep{DEVENISH2012152,
  FOLCH2012165, HEINOLD2012195, winker2012caliop,
  vernier2013advanced}.  The model outputs assuming the atmosphere has
multiple turbulent layers support the working hypothesis regarding the
effects of an atmosphere layered with respect to turbulence intensity.
Thus, in addition to the near-vent, volcanic and sedimentation
processes causing formation of volcanic layers, processes associated
with atmospheric layered turbulence also produce layering in ash
clouds.  These are often multilayered due to multiple, alternating
layers of turbulent and quiescent air.  The distal ash layers
furthermore scale to the depth of the alternating turbulent and
quiescent layers in the atmosphere, which are $\mathcal{O}[0.1-5]$ km
deep.

Because of the alternating turbulent and quiescent structure in the
troposphere and stratosphere, volcanic ash clouds tend to separate
vertically over time, lending to them the distinct layering or banded
appearance in imagery. The turbulent layers retain particles longer
than the quiescent layers because the turbulence retains particles in
suspension.  Particles fall more rapidly through the relatively
quiescent layers (lower $\kappa_z$) by single particle settling, or
because of convective sedimentation.

The results suggest that to better model the position and morphology
of ash clouds for aviation safety and other purposes in VATDs, the
vertical characteristics of the atmosphere need to be better resolved
than is typical at present.  Because of the importance of turbulence
and moisture to layer formation, it is critical that these two
parameters especially be estimated well, and at as high a vertical
resolution as possible.

\paragraph{Abbreviations.} The following abbreviations are used in this manuscript:\\

\noindent 
\begin{tabular}{@{}ll}
  AERONET & AErosol RObotic NETwork\\
  BT & Brightness Temperature\\
  CALIOP & Cloud-Aerosol Lidar with Orthogonal Polarization\\
  EARLINET & European Aerosol Research Lidar Network\\
  IAVW & International Airways Volcano Watch\\
  RH & Relative Humidity\\
  UTLS & Upper Troposphere - Lower Stratosphere\\
  VATD & Volcanic Ash Transport and Dispersal
\end{tabular}





\bibliographystyle{agu}
\bibliography{ash_strata_clouds}

\end{document}